\documentclass[a4paper]{scrartcl}

\usepackage[utf8]{inputenc}
\usepackage{latexsym}
\usepackage{amsmath}
\usepackage{amsthm}
\usepackage{amssymb}
\usepackage{mathtools}
\usepackage{stmaryrd}
\usepackage{enumitem}
\usepackage{hyperref}
\usepackage{colortbl}
\usepackage{ifthen}
\usepackage{calc}
\usepackage{xcolor}
\usepackage{authblk}
\usepackage{caption}
\usepackage{subcaption}

\newtheoremstyle{mythm}{}{}{\itshape}{}{\bfseries}{.}{.5em}{\thmname{#1}~\thmnumber{#2}\ifthenelse{\equal{\thmnote{#3}}{}}{}{~(\thmnote{#3})}}

\newtheoremstyle{mydefn}{}{}{\upshape}{}{\bfseries}{.}{.5em}{\thmname{#1}~\thmnumber{#2}\ifthenelse{\equal{\thmnote{#3}}{}}{}{~(\thmnote{#3})}}

\newtheoremstyle{myremark}{}{}{\upshape}{}{\itshape}{.}{.5em}{\thmname{#1}~\thmnumber{#2}\ifthenelse{\equal{\thmnote{#3}}{}}{}{~(\thmnote{#3})}}

\theoremstyle{mythm}
\newtheorem{theorem}{Theorem}[section]

\theoremstyle{mydefn}
\newtheorem{definition}[theorem]{Definition}
\newtheorem{example}[theorem]{Example}
\theoremstyle{myremark}

\theoremstyle{mythm}

\setlist{noitemsep,topsep=0.3em}
\setlist[enumerate,1]{label = (\arabic*)}

\newcommand{\uend}{\mbox{}\hfill$\lrcorner$}

\renewcommand{\bar}{\overline}

\newcommand{\lmulti}{{\{\hspace{-3.4pt}\{}}
\newcommand{\rmulti}{{\}\hspace{-3.4pt}\}}}

\renewcommand{\phi}{\varphi}
\renewcommand{\epsilon}{\varepsilon}

\newcommand{\Nat}{{\mathbb N}}
\newcommand{\Real}{{\mathbb R}}

\newcommand{\Rat}{{\mathbb Q}}

\newcommand{\CB}{{\mathcal B}}

\newcommand{\CD}{{\mathcal D}}

\newcommand{\CG}{{\mathcal G}}

\newcommand{\CN}{{\mathcal N}}

\newcommand{\CP}{{\mathcal P}}

\newcommand{\BD}{{\boldsymbol D}}

\newcommand{\KA}{{\mathfrak A}}
\newcommand{\KB}{{\mathfrak B}}
\newcommand{\KC}{{\mathfrak C}}

\newcommand{\KO}{{\mathfrak O}}
\newcommand{\KS}{{\mathfrak S}}

\newcommand{\MD}{{\mathbb D}}
\newcommand{\MF}{{\mathbb F}}
\newcommand{\MU}{{\mathbb U}}
\newcommand{\MX}{{\mathbb X}}

\newcommand{\DB}{{\mathbb{DB}}}
\newcommand{\DBB}{{\mathbb{DB}}^{\textup{bag}}}

\begin{document}
\title{Probabilistic Data with Continuous Distributions\thanks{The results presented in this paper were originally
    published in M.~Grohe and P.~Lindner: \emph{Infinite Probabilistic
      Databases}, Proc.~ICDT~2020 and M.~Grohe, B.L.~Kaminski,
    J.-P.~Katoen, P.~Lindner: \emph{Generative Datalog with Continuous
      Distributions}, Proc.~PODS~2020.}
}
\author[1]{Martin Grohe}
\author[2]{Benjamin Lucien Kaminski}
\author[1]{Joost-Pieter Katoen}
\author[1]{Peter Lindner}

\affil[1]{RWTH Aachen University, Germany}
\affil[2]{University College London, UK}
\date{}

\maketitle

\begin{abstract}
  Statistical models of real world data typically involve continuous
  probability distributions such as normal, Laplace, or exponential
  distributions. Such distributions are supported by many probabilistic
  modelling formalisms, including probabilistic database
  systems. 
  Yet, the traditional theoretical framework of probabilistic
  databases focuses entirely on finite probabilistic databases.

  Only recently, we set out to develop the mathematical theory of
  infinite probabilistic databases. The present paper is an exposition
  of two recent papers which are cornerstones of this theory. In
  (Grohe, Lindner; ICDT 2020) we propose a very general framework for
  probabilistic databases, possibly involving continuous probability
  distributions, and show that queries have a well-defined semantics
  in this framework. In (Grohe, Kaminski, Katoen, Lindner; PODS 2020)
  we extend the declarative probabilistic programming language
  Generative Datalog,
  proposed by (Bárány et al.~2017) for discrete probability
  distributions, to continuous probability distributions and show that
  such programs yield generative models of continuous
  probabilistic databases.  
\end{abstract}

\section{Introduction}
Probabilistic databases \cite{suc20,Suciu+2011,VandenBroeckSuciu2017} provide a framework for quantitatively
modelling uncertainty in data. Sources of uncertainty are numerous; 
common examples are noisy sensor data, data gathered from
unreliable sources, and inconsistent data. Formally, a probabilistic
database (PDB) is a
probability space over database instances, called the possible
worlds. Traditionally, these probability spaces were limited to be
finite. This implies a closed world assumption where only finitely
many facts could possibly be true, and it rules out any probability
distributions with an infinite support. 
Yet, in many applications,
infinitely, even uncountably infinitely, supported probability distributions arise
naturally, and many real-world statistical phenomena follow infinite
probability distributions such as Poisson distributions, normal
distributions, or exponential distributions.

\begin{example}\label{exa:temp}
  Suppose we have a relation storing room temperatures, as in
  Figure~\ref{fig:temp}(a). The temperature measurements may be noisy,
  and we may account for this by adding a normally distributed error
  with a small variance $\epsilon>0$, resulting in a simple
  probabilistic relation which may be represented as in
  Figure~\ref{fig:temp}(b).
  \uend
\end{example}

\begin{example}\label{exa:particles}
  In this example, consider a particle detector such as the Alpha
  Magnetic Spectrometer (AMS-02\footnote{See
    \url{https://ams02.space/}.}) on the ISS. Suppose we record the
  detected particles in a relation of schema
  $(\texttt{Time}, \texttt{Trajectory}, \texttt{Velocity})$.  As in
  the previous example, the measurements (of the trajectory and
  velocity) may be imprecise and best modelled by a probability
  distribution. But here we have an additional source of uncertainty:
  some particles may go undetected. If we also model this type of
  error, the number of tuples in the relation becomes a random
  variable as well. Then
  there is no a-priori bound on the size of the instances in the
  resulting PDB. Note, however, that every instance is still finite,
  because in every time interval only finitely many particles can hit
  the detector, and our model should account for that.  \uend
\end{example}

\begin{figure*}
  \centering
  \renewcommand{\arraystretch}{1.2}\begin{subfigure}{6.5cm}
    \centering
    {\small\ttfamily
      \begin{tabular}{|c|c|c|}
        \hline\rowcolor{lightgray}
        \texttt{RoomNo}&\texttt{Time}&\texttt{°C}\\
        \hline
        4108&2021-01-05 08:00&20.2\\
        \hline\rowcolor{lightgray}
        4108&2021-01-05 14:00&21.8\\
        \hline
        4109&2021-01-05 08:00&22.1\\
        \hline
        \vdots&\vdots&\vdots
      \end{tabular}
    }
    \caption{}
  \end{subfigure}
  \hfill
  \begin{subfigure}{7.5cm}
    \centering
    {\small\ttfamily \begin{tabular}{|c|c|c|}
        \hline\rowcolor{lightgray}
        \texttt{RoomNo}&\texttt{Time}&\texttt{°C}\\
        \hline
        4108&2021-01-05 08:00&$\CN(20.2,\epsilon)$\\
        \hline\rowcolor{lightgray}
        4108&2021-01-05 14:00&$\CN(21.8,\epsilon)$\\
        \hline
        4109&2021-01-05 08:00&$\CN(22.1,\epsilon)$\\
        \hline
        \vdots&\vdots&\vdots
      \end{tabular}\\[1ex]
    }
    \caption{}
  \end{subfigure}
    
  \caption{A relation storing room temperatures: (a)
    deterministically and (b) with a normally distributed
    error}
  \label{fig:temp}
\end{figure*}

Both examples exhibit probabilistic databases with continuous
probability distributions that cannot be captured by the traditional
model of finite probabilistic databases. Generalising from finite to
continuous probability distributions comes with a substantial
mathematical overhead. While PDBs of fixed (or bounded) size, such as
those arising from Example~\ref{exa:temp}, are still relatively easy
to handle, PDBs of unbounded size such as the one we saw in
Example~\ref{exa:particles} are nontrivial to capture mathematically,
let alone to deal with algorithmically. 
Several PDB systems that have been proposed over the years \cite{Agrawal+2009,
Jampani+2011,KennedyKoch2010,Singh+2008} handle continuous
probability distributions. The flexibility of these systems reaches as far as
providing declarative representations of continuous probabilistic databases and
even continuous-space database-valued Markov processes. Yet,
only recently~\cite{GroheLindner2019,GroheLindner2020}, we proposed a general
framework for rigorously dealing with probabilistic databases over
continuous domains and provided a sound semantics for standard query languages
such as the relational calculus. We will present this framework in
Sections~\ref{sec:standard} and \ref{sec:queries} of this paper. To
distinguish them from the traditional ``finite'' PDBs, we call PDBs
with an infinite sample space \emph{infinite PDBs} in the
following. Note that every instance in an infinite PDB is just a
standard finite relational database instance, it is only the
sample space of all possible instances that is infinite.

A difficult issue when dealing with PDBs is how to efficiently
represent them. This problem already arises
for finite PDBs, but is much more
fundamental when dealing with infinite PDBs that do not even allow
for a naive representation that explicitly lists all instances. So we
have to rely on \emph{implicit} representations, which can either be ad-hoc
representations such as the one chosen to illustrate
Example~\ref{exa:temp} in Figure~\ref{fig:temp}(b) or generic
formalisms for representing complex probability distributions, such as
probabilistic graphical models, deep neural networks, and probabilistic
programming languages. Yet, when dealing with (relational) PDBs, it is desirable to stay within the declarative
framework of relational databases. To this end, Bárány, ten Cate, Kimelfeld, Olteanu, and Vagena
\cite{Barany+2017} introduced a declarative probabilistic programming
language based on Datalog,
which has a generative part enabling to represent complex
probability distributions strictly within the framework of relational
databases. However, the semantics of Bárány et al.\ 
is only able to handle discrete probability distributions. 
In~\cite{Grohe+2020}, we generalised the semantics to continuous
distributions. The resulting \emph{Generative Datalog} can serve
both as a powerful representation language for relational PDBs with
discrete and continuous distributions and as a query language for
PDBs. We present this language in Section~\ref{sec:datalog}.

The reader may wonder if it is really necessary to consider continuous
probabilistic databases. After all, they can only be mathematical
abstractions of real systems, where instead of the continuum of real
numbers we only see the finite set of 64 bit floating point
numbers. Then aren't finite probabilistic databases all we need?  Well,
the history of computer science has shown us that the right
abstractions can be extremely powerful---just think of the relational
database model---
and certainly we do not want
the semantics of our query languages depend on whether we use 32 or 64
bit floating-point numbers to specify probabilities. All of
applied mathematics, including statistics, uses the real numbers as the
right abstraction to reason about continuous phenomena. And when
reasoning about uncertain and probabilistic data, we want to have
standard tools such as normal distributions at our
disposal.

\section{Towards Infinite PDBs}
\label{sec:intuition}

Before we delve into the mathematical details, in this
section we describe the general approach on an intuitive level and
highlight the technical difficulties we are facing.

We define a probabilistic database to be a
probability space whose sample space consists of database instances of
some schema $\tau$. In the traditional approach, this probability
space is assumed to be finite; here, we would like to allow it to be
infinite. The difficulty when defining probabilities on uncountable
spaces such as the reals is that we cannot assign a well-defined
probability to all subsets of the space, but only to subsets
that are \emph{measurable}.

Let us ignore this issue for a moment (though it will come back to
bite us) and think about \emph{how we can actually define a probability
distribution on uncountable sets of database instances}. Let us fix a
simple database schema $\tau$ consisting of a single binary relation
$R$ of schema $(\texttt{Time}, \texttt{Value})$, where the attribute
\texttt{Value} is real-valued. Instances are relations of this schema.
We can also view them as finite sets (without duplicates) or finite bags (possibly with duplicates)---depen{\-}ding on the
type of semantics we are interested in---of facts of the form $R(t,v)$,
where $t$ is a point in time and $v$ a real number. If we want to
define a finite probability space on the instances, we can simply pick
a finite set $D_1,\ldots,D_m$ of instances and assign probabilities
$p_1,\ldots,p_m\in[0,1]$ to them such that $\sum_ip_i=1$. We
can extend this approach to countably infinite spaces, but not to
uncountable spaces, where typically every single instance has
probability $0$. This happens, for example, if we assume the
\texttt{Value} to be normally distributed at any \texttt{Time}. We
know how to define a probability distribution on the \texttt{Value}s
(that is, the real numbers); we only need to specify the probability
mass on each interval. But here we need to define a probability
distribution on \emph{sets} or \emph{bags} of
\texttt{Time}-\texttt{Value} pairs. It is not at all obvious how to do
that, except maybe in simple settings such as the one described in
Example~\ref{exa:temp}. We need to draw from the theory of
\emph{finite point processes}~\cite{Moyal1962,Macchi1975,DaleyVere-Jones2003}. In probability theory, point processes
are used to describe probability spaces of finite or countable sets or
bags. Based on the theory of point processes, we will define
a very general framework for infinite PDBs that we call \emph{standard
  PDBs} (see Section~\ref{sec:standard}).

Once we have defined our probability spaces, we need to think about
\emph{querying PDBs}. To define the semantics of queries and views, let us
consider a view $V$ mapping instances of schema $\tau$ to instances
of schema $\tau'$. Queries are just specific views where the target
schema $\tau'$ consist of a single relation schema. We want to define
a semantics for this view $V$ on probabilistic databases, that is, we
want to extend it to a mapping from PDBs of schema $\tau$ to PDBs of
schema~$\tau'$. Let us assume that we have a PDB $\CD$ of schema~$\tau$, 
and we want to define the image $V(\CD)$, which is supposed
to be a PDB of schema~$\tau'$. 
To do this, for a set~$\BD'$ of
instances of schema~$\tau'$ we define the probability of $\BD'$ in~$V(\CD)$ to be
the probability of the set~$V^{-1}(\BD')$ in $\CD$:
\begin{equation}
  \label{eq:1}
  \Pr_{V(\CD)}(\BD')\coloneqq\Pr_{\CD}\big(V^{-1}(\BD')\big). 
\end{equation}

\begin{example}\label{exa:temp2}
  Recall Example~\ref{exa:temp}, where we considered PDBs of a schema
  \[\tau=\{\texttt{Temp}(\texttt{RoomNo},\texttt{Time},\texttt{°C})\}.\] Entries
  are room temperatures at various times. Let $Q$ be the query that
  maps instances of schema $\tau$ to instances of schema $\tau'=\{\texttt{AvTemp}(\texttt{RoomNo},\texttt{°C})\}$ recording
  the average temperature in each room, defined by the SQL-expression
  \begin{center}
    \texttt{SELECT RoomNo, AVG(°C) FROM Temp GROUP BY RoomNo}.
  \end{center}
  Let us apply this query to the PDB $\CD$ represented by
  the relation \texttt{Temp} shown in Figure~\ref{fig:temp2}.
  \begin{figure}
    \centering
    {\ttfamily
    \begin{tabular}{|c|c|c|}
\hline\rowcolor{lightgray}
      \texttt{RoomNo}&\texttt{Time}&\texttt{°C}\\
      \hline
      4108&2021-01-05 08:00&$\CN(20.2,0.1)$\\
      \hline\rowcolor{lightgray}
      4108&2021-01-05 14:00&$\CN(21.8,0.1)$\\
        \hline
      4109&2021-01-05 08:00&$\CN(22.1,0.1)$\\
      \hline\rowcolor{lightgray}
      4109&2021-01-05 14:00&$\CN(22.3,0.1)$\\
        \hline
      4109&2021-01-06 08:00&$\CN(21.9,0.1)$\\
      \hline
    \end{tabular}}
    \caption{A PDB of schema
      $\tau=\{\texttt{Temp}(\texttt{RoomNo},\texttt{Time},\texttt{°C})\}$}
    \label{fig:temp2}
  \end{figure}
  Note that in all instances of this PDB, the table \texttt{Temp} has
  exactly five rows recording the temperatures in room 4108 at two
  different times and the temperatures in room 4109 at three
  different times. For simplicity, we assume that the random variables 
  describing the entries in the five rows are independent.

  In every instance of $Q(\CD)$, the table \texttt{AvTemp} has exactly
  two rows recording the average temperatures in rooms 4108 and
  4109. We can easily compute the probabilities in $Q(\CD)$. For
  example, the probability that both rooms have an average
  temperature in the range 20--22 degrees equals the probability
  that the average of two normally distributed random variables with
  means $20.2,21.8$ and variance $0.1$ is between $20$ and
  $22$ times the probability that the average of three normally
  distributed random variables with means $22.1,22.3,21.9$ and
  variance $0.1$ is between $20$ and $22$. Actually,
  the table \texttt{AvTemp} in $Q(\CD)$ can be represented as follows.
   \begin{center}
    \begin{tabular}{|c|c|c|}
\hline\rowcolor{lightgray}
      \texttt{RoomNo}&\texttt{°C}\\
      \hline
      4108&$\CN(21.0,0.05)$\\
      \hline\rowcolor{lightgray}
      4109&$\CN(22.1,0.033)$\\
      \hline
    \end{tabular}
  \end{center}
   The fact that a linear
  combination of normal distributions is again a normal distribution
  enables us to represent $Q(\CD)$ in such a simple ``closed
  form''. In general, views of PDBs can be far more complicated than
  \mbox{the original PDBs.} ~\uend
\end{example}

Unfortunately, there is a subtle issue that we have neglected when
defining the semantics of views and queries over PDBs. Recall
that in uncountable probability spaces, we cannot define probabilities
for all subsets of the sample space, but only for so-called \emph{measurable}
sets. This means that in the definition \eqref{eq:1}, we only need to
consider measurable sets $\BD'$ of instances of schema $\tau'$, but
we need to make sure that the set $V^{-1}(\BD')$ is
measurable as well, for otherwise the probability on the
right-hand side of \eqref{eq:1} is not defined. This means that a view
$V$ only has a well-defined semantics on probabilistic databases if
for every measurable set $\BD'$ in the target space the pre-image
$V^{-1}(\BD')$ is a measurable set in the source space. If this is the
case, we call $V$ measurable. \emph{Only measurable views and
  queries have a well-defined semantics on probabilistic databases.}
Fortunately, it turns out that all views defined in standard query
languages such as the relational calculus or Datalog are
measurable. But this is a nontrivial result
(Theorem~\ref{theo:standardPDB}). In \cite[Example~8]{GroheLindner2020}, we give an
example of a relatively simple ``query'' that is not measurable.

\section{Mathematical Background}

In this section, we collect some mathematical background
underlying our approach to PDBs. The reader may skip this
section and use it as a \mbox{reference whenever needed later}.

\subsection*{Topology}
Topology qualitatively captures concepts such as closeness,
convergence, and continuity, and it is the foundation for the measure
theory and probability theory we need here. A \emph{topology} on a set
$\MX$ is a family $\KO$ of subsets of $\MX$ that contains~$\MX$ and
the empty set and is closed under arbitrary unions and finite intersections. We
call $(\MX,\KO)$ a \emph{topological space} and the elements $O\in\KO$
\emph{open sets}.

\begin{example}\label{exa:top}
  \begin{enumerate}
  \item In the \emph{standard topology} on the reals $\Real$, a set
    $O\subseteq \Real$ is open if for every $x\in O$ there is an
    $\epsilon>0$ such that $(x-\epsilon,x+\epsilon)\subseteq O$. Note
    that this topology is \emph{generated} by the open intervals,
    which means that every open set is the union of open
    intervals.\footnote{We take the union over the empty family
      of sets to be the empty set.}
  \item For every set $\MX$, the power set $2^{\MX}$ is a topology on
    $\MX$, the so-called \emph{discrete topology}.\uend
  \end{enumerate}
\end{example}

For $i=1,2$, let $(\MX_i,\KO_i)$ be a topological space.
A function $f\colon\MX_1\to\MX_2$ is \emph{continuous} (with respect to $\KO_1,\KO_2$) if
$f^{-1}(O_2)\in\KO_1$ for every $O_2\in\KO_2$. 

Every \emph{metric} $d$ on $\MX$ (that is, a distance function on pairs
of elements of $\MX$ that is symmetric, satisfies the triangle
inequality, and has the property that two points have distance~$0$ if
and only if they are equal) induces a topology on $\MX$ where a set
$O\subseteq \MX$ is open if for every $x\in O$ there is an
$\epsilon>0$ such that $\{y\mid d(x,y)<\epsilon\}\subseteq O$. A
topological space $(\MX,\KO)$ is \emph{metrisable} if it is induced by a
metric on $\MX$ in this way. Obviously, the standard topology on the
reals (Example~\ref{exa:top}(1)) is metrisable. The discrete topology
on an arbitrary set $\MX$ (Example~\ref{exa:top}(2)) is metrisable as well; as a metric we use
the function $d$ with $d(x,x)=0$ and $d(x,y)=1$ for all $x\neq y$, also know as the \emph{discrete metric}.

A topological space $(\MX,\KO)$ is \emph{separable} if there is
a countable subset $Y\subseteq\MX$ such that every nonempty open set
$O\in\KO\setminus\{\emptyset\}$ contains an element from $Y$ (we say
that $Y$ is \emph{dense}). For
example, for the reals with the standard topology, the
set $\Rat$ of rationals is a dense subset. The discrete topology on a
set $\MX$ is separable if and only if $\MX$ is countable. Separability
is a very important technical property in our arguments, because it
enables us to work with countable approximations.

A final condition we need (though it is less important for us) is
completeness: intuitively, a metrisable topological space $(\MX,\KO)$
is \emph{complete} if every convergent sequence (more precisely,
Cauchy sequence) converges to a point in $\MX$. We omit the formal
definition. A \emph{Polish space}  is a complete, separable, metrisable
topological space (and its topology is \emph{Polish}). The reals with the standard topology, all finite-dimensional Euclidean spaces, and all countable discrete topological
spaces are Polish spaces. 

\emph{It is safe to say that all topological
spaces we will ever find in database applications are Polish spaces.}

\subsection*{Measure Theory and Probability}

A \emph{$\sigma$-algebra} on a set $\MX$ is a set $\KA$ of
subsets of $\MX$ that contains the empty set and is closed under
complementation and countable unions. A pair
$(\MX,\KA)$, where $\KA$ is a $\sigma$-algebra on $\MX$, is called a
\emph{measurable space}.
\begin{example}\label{exa:sigma}
  \begin{enumerate}
  \item For every set $\MX$, the set $\{\emptyset,\MX\}$ and the power
    set $2^{\MX}$ are    $\sigma$-algebras on $\MX$.

  \item
    Another $\sigma$-algebra on $\MX$ is
    the set of all $Y\subseteq \MX$ such that either $Y$ is countable
    or $\MX\setminus Y$ is countable.
  \item The set of all Lebesgue measurable subsets of the reals is a
    $\sigma$-algebra.\uend
  \end{enumerate}
\end{example}

Let $\MX$ be a set and $\KS\subseteq 2^{\MX}$. The $\sigma$-algebra
\emph{generated} by $\KS$ is the closure of $\KS$ under complementation and
countable intersections, that is, the smallest $\sigma$-algebra on
$\MX$ that contains $\KS$. Observe that the $\sigma$-algebra defined
in Example~\ref{exa:sigma}(2) is the $\sigma$-algebra generated by all
singleton sets $\{x\}$ for $x\in\MX$.

For any topological space $(X,\KO)$, the $\sigma$-algebra generated by the
topology $\KO$ is called the \emph{Borel $\sigma$-algebra} on $\MX$, and its elements are called \emph{Borel sets}. A
measurable space $(\MX,\KA)$ is a \emph{standard Borel space} if $\KA$
is the Borel $\sigma$-algebra of some Polish topology on $\MX$. It is
not difficult to show that if $d$ is a metric inducing such a Polish
topology and $Y$ is a countable dense subset then $\KA$ is generated
by the countable set of open balls
$B_{1/n}(y)\coloneqq \{x\in\MX\mid d(x,y)<1/n\}$ for positive integers
$n$ and $y\in Y$. This is one of the reasons making standard Borel
spaces very convenient to handle.

For $i=1,2$, let $(\MX_i,\KA_i)$ be a measurable space.  A function
$f\colon\MX_1\to\MX_2$ is \emph{measurable} (with respect to $\KA_1,\KA_2$)
if $f^{-1}(A_2)\in\KA_1$ for every $A_2\in\KA_2$. If $\KA_i$ is the
Borel $\sigma$-algebra of some topology $\KO_i$ on $\MX_i$, then every
continuous function is measurable; the converse does not always hold.
The \emph{Cartesian product}
of $(\MX_1,\KA_1)$ and $(\MX_2,\KA_2)$ is
the measurable space $(X_1\times X_2,\KA_1\otimes\KA_2)$, where
$\KA_1\otimes\KA_2$ is the $\sigma$-algebra generated by the sets
$A_1\times A_2$ for $A_i\in\KA_i$. If $\MX_1$ and $\MX_2$ are
disjoint, then the (disjoint) union of the two measurable spaces is
the measurable space $(X_1\cup X_2,\KA_1\oplus\KA_2)$, where
$\KA_1\oplus\KA_2$ is the set of all sets $A\subseteq X_1\cup X_2$
such that $A\cap X_i\in\KA_i$. It can be shown that if the spaces
$(\MX_i,\KA_i)$ are standard Borel spaces then
$(X_1\times X_2,\KA_1\otimes\KA_2)$ and
$(X_1\cup X_2,\KA_1\oplus\KA_2)$ are standard Borel spaces as well.

Let $(\MX,\KA)$ be a measurable space. A \emph{measure} on $(\MX,\KA)$
is a function $M$ from $\KA$ to $\Real_{\ge 0}\cup\{\infty\}$ (the
nonnegative reals extended by infinity) that is
\emph{$\sigma$-additive}, that is, for every countable family
$A_1,A_2,\ldots$ of mutually disjoint sets in $\KA$ it holds that
$M(\bigcup_{i\ge 1}A_i)=\sum_{i\ge 1}M(A_i)$. A measure $M$ is
\emph{finite} if $M(\MX)<\infty$, and it is a
\emph{probability measure} (or a \emph{probability distribution}) if $M(\MX)=1$. We call $(\MX,\KA,M)$ a
\emph{measure space}, or a \emph{probability space} if $M$ is a
probability measure. $\MX$ is called the \emph{sample space} and $\KA$
the \emph{event space} of this probability space.

\begin{example}\label{exa:probspaces}
  \begin{enumerate}
  \item Let $(\MX,\KA)$ be a measurable space and $Y\subseteq \MX$. We
    define a measure $M$ by letting $M(A)$ be the cardinality of
    $|A\cap Y|$ (either finite or $\infty$). $M$ is what we call a
    \emph{counting measure}. It is finite if $Y$ is finite.
  \item The \emph{normal distribution} $\CN(\mu,\sigma)$ is the unique
    probability measure $P$ on the standard Borel space $(\Real,\KB)$
    with
    \[
      P\big((a,b]\big)=\frac{1}{\sigma\sqrt{2\pi}}\int_a^b
      \exp\left(-\frac{(x-\mu)^2}{2\sigma^2}\right)dx.
      \]
      for all $a<b$. (It can be shown that a probability measure on the
      Borel $\sigma$-algebra over the reals is determined by its
      values on the half open intervals.)
    \item
      \sloppy
      If $\MX$ is a countable set (finite or countably infinite),
      then we can define a probability measure $P$ on $(\MX,2^{\MX})$ by
      defining the singleton values $P(\{x\})\in[0,1]$ such that 
      $\sum_{x\in X}P(\{x\})=1$ and letting $P(A)\coloneqq\sum_{x\in A}P(\{x\})$ for all $A\subseteq\MX$. In
      fact, every probability measure on $\MX$ can be defined this
      way, regardless of what the $\sigma$-algebra is. So for
      countable probability spaces, we can always assume that the
      $\sigma$-algebra is the power set of the sample space $\MX$.
      \uend
  \end{enumerate}
\end{example}
\section{Standard PDBs}
\label{sec:standard}
Let $\tau$ be a database schema, and let $\MU_\tau$, the 
\emph{universe}, be the union of the domains of all attributes
occurring in $\tau$. We view \emph{database instances} as finite sets
or bags (a.k.a.~multisets) of \emph{facts} of the form
$R(a_1,\ldots,a_k)$, where $R(A_1,\ldots,A_k)$ is a relation schema in
$\tau$ and for every $i$ the value $a_i\in\MU_\tau$ is contained in
the domain of attributes $A_i$. Even if we are only interested in
set instances, for technical reasons we need to consider bag instances
as well. We denote the set of all facts over $\tau$ by $\MF_\tau$ and
the set of database instances over $\tau$, that is, finite bags of
facts in $\MF_\tau$, by $\DBB_\tau$. Moreover, we denote the subset of
all plain sets of facts, that is, set instances, by $\DB_\tau$. In all
these notations, we omit the subscript ${}_\tau$ if the schema is
clear from the context or irrelevant.

Under the most general definition, a \emph{probabilistic database} is
just a probability space $\CD=(\MD,\KA,P)$, where
$\MD\subseteq\DBB$. However, it is very difficult to work with this
general definition. While we may have an intuition about defining the
sample space and the probabilities, it is completely unclear how to
define the event space, that is, the $\sigma$-algebra $\KA$, which is
a set of sets of bags of facts (sic).

\begin{example}
  Recall Examples~\ref{exa:temp}, \ref{exa:temp2}, and let
  $\CD=(\MD,\KA,P)$ be the (informally described) PDB shown
  in Figure~\ref{fig:temp2}. The sample space $\MD\subseteq\DB_\tau$
  consists of
  all instances
  \begin{align*}
    \big\{&\texttt{Temp(4108,2021-01-05 08:00,$t_1$)},\\
    &\texttt{Temp(4108,2021-01-05 14:00,$t_2$)},\\
    &\texttt{Temp(4109,2021-01-05 08:00,$t_3$)},\\
    &\texttt{Temp(4109,2021-01-05 14:00,$t_4$)},\\
    &\texttt{Temp(4109,2021-01-06 08:00,$t_5$)}
    \big\},
  \end{align*}
  where $t_1,\ldots,t_5\in\Real$. We have seen in
  Example~\ref{exa:temp2} how to calculate the probability of a set of
  instances. However, it is not obvious which sets are measurable,
  that is, have a
  well-defined probability and therefore should belong to the
  $\sigma$-algebra $\KA$. Intuitively, at least sets such as those
  considered in Example~\ref{exa:temp2} where the temperatures are in
  certain intervals, should be measurable.

  In this simple example, we can define a suitable $\sigma$-algebra by
  an ad-hoc
  product construction starting from the Borel $\sigma$-algebra on the
  reals, but already in the only slightly more complicated setting of
  Example~\ref{exa:particles}, where the number of tuples in an
  instance is also a random variable that is a-priori unbounded, it
  becomes difficult to carry out such \mbox{a construction}.
  \uend
\end{example}

The point is: even if we can somehow come up with an ad-hoc
construction of a $\sigma$-algebra for every PDB that we want
to work with, reasoning about $\sigma$-algebras is definitely not what
we want to do when working with probabilistic data. Yet, as we have seen
in Section~\ref{sec:intuition}, measurability is an issue when giving queries and views a meaningful semantics.

A solution to this dilemma is a theoretical framework that gives us a
\emph{generic} construction of $\sigma$-algebras only depending on the
schema $\tau$ and the universe $\MU$ that is rich enough to make all
sets that we typically want to consider measurable and at the same
time ensures that all reasonable queries and views are
measurable. \emph{Standard probabilistic databases}, introduced in
\cite{GroheLindner2020}, provide such a framework.

The main technical challenge is to generically construct a
sufficiently rich $\sigma$-algebra $\KA=\KA_\tau$ on $\DBB$. If the
universe $\MU$ is countable, then the set $\MF$ of facts and hence the
set $\DBB$ of all finite bags of facts are countable as well, and we
can simply let $\KA=2^{\DBB}$ (see Example~\ref{exa:probspaces}(3)).
But what do we do if the universe is uncountable? The additional
assumption we need to make is that we have \emph{a topology on the universe,
in fact a Polish topology}. As uncountable universes we may see in
typical database applications are usually derived from the reals in
some way, this is \mbox{no serious restriction}.

\begin{example}
  Typical domains of database attributes are integers, reals, strings,
  and time stamps. So we will have a universe like
  $\MU=\Sigma^*\cup\Real$ for some finite alphabet $\Sigma$ (say,
  UTF8). If by $\KO_{\Real}$ we denote
  the standard topology on the reals,
  the generic way of extending it to a Polish topology $\KO$ on
  $\MU$ is to let $\KO$ be the set of all
  $O\subseteq \MU$ such that $O\cap\Real\in\KO_{\Real}$. It is
  straightforward to extend this construction to more complicated
  universes where we add, for example, a set of (uncountably many) time
  stamps. 
  \uend
\end{example}

Let us assume in the following that $\KO_{\MU}$ is a Polish topology
on the universe $\MU$. We assume that this topology is part of the
information provided by the schema $\tau$.
Let $\KA_{\MU}$ be the $\sigma$-algebra
generated by $\KO_{\MU}$. Then $(\MU,\KA_{\MU})$ is a standard Borel
space. Using finite Cartesian products and disjoint
unions, we can lift $\KA_{\MU}$ to a $\sigma$-algebra $\KA_{\MF}$ on
the set $\MF$ of facts. $(\MF,\KA_{\MF})$ is still a standard Borel
space.

The next step will be to lift $\KA_{\MF}$ to a $\sigma$-algebra $\KC$
on $\DBB$. Maybe the most direct way of doing this is to first lift
the $\sigma$-algebra to all finite tuples of facts using finite
Cartesian products and a countable disjoint union and then ``factor''
the resulting $\sigma$-algebra through all permutations to go from
tuples to bags. A more elegant way of defining the same
$\sigma$-algebra is as follows.\footnote{It is not obvious that the
  two constructions indeed lead to the same $\sigma$-algebra. This
  follows from a theorem from point-process theory.} For every set $F\subseteq\MF$ of facts and
every instance $D\in\DBB$, we let $|D|_{F}$ be the number of elements
of $F$ in $D$ counted according to their multiplicities. For example,
$|\lmulti f,f,g,g,g, h\rmulti|_{\{f,g\}}=5$. For $n\in\Nat$, we let
$\#(F,n)$ be the set of all $D\in\DBB$ with $|D|_F=n$. Finally, we let
$\KC\coloneqq\KC_\tau$ be the $\sigma$-algebra generated by all sets
$\#(F,n)$ for $F\in\KC_{\MF}$ and $n\in\Nat$. Since $\KC$ is generated
by the \emph{counting events} $\#(F,n)$, it is called the
\emph{counting $\sigma$-algebra} (hence the letter $\KC={}$``Fraktur
C''). 
Another way of seeing $\KC$ is that it is the smallest $\sigma$-algebra such that for all
measurable sets $F\in\KA_{\MF}$ of facts, the function
$|\,\cdot\,|_F\colon\DBB\to\Nat$ is measurable with respect to $\KC$ and
$2^{\Nat}$. We will see that this is enough to guarantee that all
queries defined in standard query languages are measurable as well.

\begin{definition}
  A \emph{standard probabilistic database} is a probability space
  $(\DBB_\tau,\KC_\tau,P)$ for some schema $\tau$.
\end{definition}

To keep the definition as simple as possible, we let the sample
space of a standard PDB be the set $\DBB_\tau$ of all bag
instances. As a result, every standard PDB can be specified by
its probability distribution. We could adopt a more liberal definition
where the sample space consists of an arbitrary measurable subset and then
restrict the $\sigma$-algebra to this set. That is, we could also
admit PDBs of the form
$
(\MD,\KC|_{\MD},P),
$
where $\MD\in\KC$ and
$\KC|_{\MD}\coloneqq\{ C\cap\MD\mid C\in\KC\}$. But note that this
space is essentially the same as the standard PDB $(\DBB,\KC,P')$
where $P'(C)\coloneqq P(C\cap\MD)$ for all $C\in\KC$. Therefore, it is
safe to view such PDBs with restricted sample spaces as standard PDBs. In
particular, since the set $\DB$ of all set instances is measurable,
this applies to \emph{standard set PDBs} of the form
$(\DB,\KC|_{\DB},P)$.

\section{Query Semantics}
\label{sec:queries}
A \emph{view} with \emph{input schema} $\tau$ and \emph{output schema}
$\tau'$ is a mapping $V\colon\DBB_\tau\to\DBB_{\tau'}$. A \emph{query} is a
view where the output schema consists of a single relation.
We call a view $V\colon\DBB_\tau\to\DBB_{\tau'}$ \emph{measurable} if it is
a measurable mapping with respect to $\KC_\tau$ and $\KC_{\tau'}$. Such a
measurable view $V$ can be lifted to standard PDBs as follows: for every standard
PDB $\CD=(\DBB_{\tau},\KC_{\tau},P)$, let $V(\CD)$ be the standard
PDB $(\DBB_{\tau'},\KC_{\tau'},P')$ where $P'$ is defined by 
\[
  P'(C')\coloneqq P\big(V^{-1}(C)\big)
\]
for all $C'\in\KC_{\tau'}$. Note that this is exactly semantics
defined in \eqref{eq:1}.

Thus a view has a well-defined semantics on standard PDBs if and only if
it is measurable. The following theorem, which is the main result of
\cite{GroheLindner2020}, states that this is the case for a wide class
of views.

\begin{theorem}\label{theo:standardPDB}
  All queries and views expressible in the relational calculus
  (with aggregation) and Datalog as well as variants such as
 Inflationary Datalog and Least Fixed-Point Logic (see~\textnormal{\cite{Abiteboul+1995}}) are measurable.
\end{theorem}

Let us remark that this theorem applies to both set semantics and bag
semantics.
The proof is a tedious inductive proof that goes through all operators
used to define the different query languages. The most involved steps
are basic relational-algebra operators such as Cartesian product or
projection. The following example exhibits some of the arguments in
an easy case that nevertheless already illustrates why we want our
underlying topological space to be Polish.

\begin{example}
  Let $\tau=\{R(A,B)\}$, where the attributes $A,B$ have the same
  domain $\MU$. We consider the equality query $Q$ that maps $R$ to
  its diagonal, that is, the selection
  \begin{center}
    \texttt{SELECT A,B FROM R WHERE A = B}.
  \end{center}
  Let $d$ be a metric on $\MU$ that induces the Polish topology 
  $\KO_{\MU}$ we assume to exist, and let $Y \subseteq \MU$ be a countable
  dense set.  Let $\KA_{\MU}$ be the Borel $\sigma$-algebra on $\MU$. Then
  $\KA_{\MU}$ is generated by the open balls $B_{1/n}(y)=\{x\in\MU\mid
  d(x,y)<1/n\}$ for $y\in Y$ and $n\in\Nat_{>0}$. Let 
  $\KA_{\MF}$ be the lifted $\sigma$-algebra on $\MF$, and
  $\Delta:=\{R(x,y)\in\MF\mid x=y\}$ be the diagonal selected by
  the query $Q$.

  As a first step, we need to prove that $\Delta\in\KA_{\MF}$.
  This is done by characterising its complement $\Delta^\complement$.
  We identify the space
  $\MF=\{R(x,y)\mid x,y\in\MU\}$ with the Cartesian product
  $\MU\times\MU$ and $\KA_{\MF}$ with the product $\sigma$-algebra
  $\KA_{\MU}\otimes\KA_{\MU}$ generated by the sets
  $A\times A'$ for $A,A'\in\KA_{\MU}$.
  Then, $\Delta$ becomes $\{(x,x)\mid x\in\MU\}$. 
  Observe that for $(x,x')\in\Delta^\complement$, there are $y,y'\in Y$ 
  and an $n\in\Nat_{>0}$ such that 
  $(x,x')\in B_{1/n}(y)\times B_{1/n}(y')$ and $\big(B_{1/n}(y)\times
  B_{1/n}(y')\big)\cap\Delta =\emptyset$. Thus
  \[
    \Delta^\complement= {} \qquad \qquad \bigcup_{\mathclap{\substack{y,y'\in Y\\B_{1/n}(y)\times
        B_{1/n}(y') \, \cap \, \Delta =\emptyset}}}\qquad B_{1/n}(y)\times
    B_{1/n}(y'),
  \]
  which is a countable union of sets in
  $\KA_{\MU}\otimes\KA_{\MU}$. Since every $\sigma$-algebra is closed
  under complementation and countable intersections, it follows that
  $\Delta\in \KA_{\MU}\otimes\KA_{\MU}$.

  To prove that the query $Q$, formally a mapping from $\DBB_\tau$ to
  $\DBB_\tau$, is measurable,
  we need to prove that for every $C\in\KC$ the pre-image $Q^{-1}(C)$
  is in $\KC$ as well. As the counting events $\#(F,n)$ for $n\in\Nat$
  and $F\in\KA_{\MF}$ generate $\KC$, it suffices to prove that the
  pre-image of each such counting event is in $\KC$. Observe that for
  every instance $D\in\DBB$ we have $Q(D)\in\#(F,n)$ if and only if $D$ contains exactly $n$
  facts $R(x,y)\in F$ with $x=y$ (counted according to
  multiplicity), or equivalently, $R(x,y)\in F\cap\Delta$. That is,
  \[
    Q^{-1}(\#(F,n))=\#(F\cap\Delta,n).
  \]
  $F,\Delta\in\KA_{\MF}$ imply $F\cap\Delta\in\KA_{\MF}$ and thus
  $\#(F\cap\Delta,n)\in\KC$. 
  \uend
 \end{example}

\section{Representations}
An infinite PDB viewed as a probability distribution
over database instances is an idealised mathematical concept that
allows us to give semantics to PDBs and queries. It is not something
that we can ever materialise. When designing probabilistic database
systems, we need to think about finite representations of the
probability spaces.

The most common model for finite probabilistic databases is that of
\emph{tuple-independent (TI)} PDBs. We can adopt the notion of TI PDBs
to countable PDBs; to represent a countably infinite TI PDB we only
need to represent a function that assigns a probability to every
fact. Countably infinite TI PDBs were studied in
\cite{GroheLindner2019}. An extension to uncountable PDBs, called
\emph{Poisson PDBs}, was proposed in \cite{GroheLindner2020b}.
Another basic model, \emph{block-independent disjoint (BID)} PDBs, can
also be extended to the infinite setting
\cite{GroheLindner2019,GroheLindner2020b}. Both TI and BID PDBs can
only represent very simple probability distributions. To obtain more
sophisticated distributions, we can apply transformations such as
views to such PDBs (views of countable TI and BID PDBs were studied in
\cite{cargrolinsta20}) or combine several PDBs into a new one using
constructions such as convex combinations and superpositions (see
\cite{GroheLindner2020b}).

A different and more general approach is to start from a deterministic
set of data, feed it into some generative model, and interpret the
output as a probability distribution on database instances. With this
approach, we are free to use all kinds of probabilistic modelling
formalisms, for example, database-valued Markov processes
\cite{Jampani+2011}, logical formalisms such as Markov Logic Networks
\cite{RichardsonDomingos2006} or ProbLog~\cite{DeRaedt+2007}, deep
neural network models such as variational
autoencoders~\cite{kinwel19}, or programs in some probabilistic
programming language \cite{barthe_katoen_silva_2020}. The difficulty
is to specify such models in a way that the output can be interpreted
as a meaningful probability distribution on database
instances. Generative Datalog (GDatalog), introduced in
\cite{Barany+2017,Grohe+2020}, is a declarative probabilistic
programming language that remains within the framework of relational
databases and therefore avoids this difficulty; the output of a
Generative Datalog program is a \mbox{PDB by definition}.

\section{Generative Datalog}
\label{sec:datalog}
\emph{Throughout this section, we assume a set semantics for
  relational databases.} For recursive languages like Datalog, a bag
semantics is less convenient, because we want to avoid repeatedly
generating new copies of the same fact.

We start by informally reviewing Datalog (see \cite{Abiteboul+1995}
for more background). A \emph{Datalog
  program} is a finite set of \emph{rules} of the form
\begin{equation}
  \label{eq:3}
  R(\bar x) \gets S_1(\bar x_1),\ldots,S_m(\bar x_m), 
\end{equation}
where
$R,S_i$ are relation symbols and $\bar x,\bar x_i$ are tuples of
variables of the appropriate lengths such that all variables in the
tuple $\bar x$ appear in one of the tuples $\bar
x_i$. The \emph{head} of the rule \eqref{eq:3} is $R(\bar x)$ and the
\emph{body} is $S_1(\bar x_1),\ldots,S_m(\bar x_m)$. The relations
appearing in the head of some rule of a Datalog program are
\emph{intensional}; all other relations are \emph{extensional}. 
The extensional (intensional) relations form the \emph{extensional}
  (\emph{intensional}, resp.) \emph{schema} of the program.

  Consider the rule \eqref{eq:3}. Given an interpretation of all
  the body relations $S_i$ and an assignment $\alpha$ to the variables
  $\bar x_i$, the rule is applicable if for all $i$ the fact
  $S_i(\alpha(\bar x_i))$ holds true
  under the current interpretation of $S_i$. The
  application of the rule generates the new fact $R(\alpha(\bar x))$.

  We run
  a Datalog program $\CP$ on a database instance over the extensional
  schema. The program iteratively computes interpretations for all the
  intensional relation symbols. All intensional relations are
  initialised to be empty. Then the rules of the program are
  applied repeatedly until no more new facts can be generated. It can
  be shown that the final relations do not
  depend on the order in which the rules are applied. Furthermore, the
  program (when applied to a finite input instance) always terminates
  in finitely many steps and hence the result can be interpreted as a
  database instance over the intensional schema. In other
  words, a Datalog program expresses a view mapping instances over the
  extensional schema to instances over the intensional schema.

  The following example illustrates these Datalog definitions and then
  develops the main ideas of its probabilistic extension
  \emph{GDatalog}.

\begin{example}\label{exa:datalog}
  Let $\CP$ be the following simple Datalog program with extensional
  relations $S,E$ and an intensional relation $R$:
  \begin{align}
    \label{eq:rule1}
    R(x)&\gets S(x),\\
    \label{eq:rule2}
    R(x)&\gets R(y), E(y,x).
  \end{align}
  We can interpret instances over the extensional schema as
  directed graphs with a distinguished set of source vertices. Then
  the program computes the set of all vertices reachable from the
  source vertices.

  Now suppose we do not only want to compute whether a vertex is
  reachable from a source, but also how long it takes to reach
  it. Assume that the edge relation is now ternary, where we
  interpret the third, real valued component as a length or traversal
  time. Consider the following program:
  \begin{align}
    \label{eq:rule3}
    R(x,0)&\gets S(x),\\
    \label{eq:rule4}
    R(x,t+s)&\gets R(y,t), E(y,x,s).
  \end{align}
  This is no longer a Datalog program in the strict sense. Yet it
  seems clear what its semantics is; after executing the program, the
  binary relation $R$ will contain all pairs $(x,t)$ such that $x$ is
  reachable from a source vertex by a walk (that is, a path with possibly
  repeated vertices and edges) of length $t$. There is however a problem with
  this: if the graph is cyclic, there may be arbitrarily long walks,
  and the output relation will no longer be finite. Therefore, let us
  assume that the input graph is acyclic.

  Now assume the traversal time of an edge is not deterministic, but
  random. Say, we model it by a log-normal distribution
  $\mathcal{LN}(\ln s,1/10)$ where the parameter $s$ is the median of this
  distribution. We may write the following program:
   \begin{align}
    \label{eq:rule5}
    R(x,0)&\gets S(x),\\
    \label{eq:rule6}
    R(x,t+\mathcal{LN}(\ln s,1/10))&\gets R(y,t), E(y,x,s).
   \end{align}
   The output of this program is supposed to be a random relation
   that contains pairs $(x,t)$,
   where $t$ is a sampled travel time along some walk from a source to
   $x$ in the input graph. We can interpret the probability space of
   all possible output relations as a PDB of schema $\{R\}$. 
   Our \emph{GDatalog program} applied to an acyclic input
   graph thus represents a PDB. 

   But now another problem with termination pops up, even if the
   input graph is acyclic. 
   The intuitive interpretation of an application of rule \eqref{eq:rule6} 
   is that for $x,t,s,y$ matching the body of the rule, we sample a value 
   $r$ from the log-normal distribution $\mathcal{LN}(\ln s,1/10)$ and then
   generate the fact $R(x,t{+}r)$. 
   But if we would apply the same rule again to the same $x,t,s$, almost surely 
   we would not sample the same $r$ again, but an $r'\neq r$, and hence 
   generate a new fact $R(x,t{+}r')$. We could do this
   over and over again and would obtain an infinite relation $R$, and
   moreover, the program would never terminate. This
   is clearly not what we want. Note that this cannot happen with a
   deterministic rule like \eqref{eq:rule4}.

   Our simple solution to avoid this problem is to stipulate that \emph{we
   can only apply the rule once to every triple $(x,t,s)$ of parameters}.
   There may, however, be application scenarios where it is desirable to
   sample from the distribution more than once. 
   To accommodate this, \emph{we allow the same rule to appear several 
   times in a program}. 
   Then for each instantiation of the rule, we can sample once. 
   A more flexible, but more complicated way of sampling several 
   times with the same parameter tuple is to introduce another parameter 
   that serves as an index for the samples.
  \uend
\end{example}

Rules \eqref{eq:rule5} and \eqref{eq:rule6} give a typical example of
a GDatalog program.  
To define GDatalog programs in general, besides the extensional 
and intensional schema we need to specify a family $\Psi$ of 
\emph{parametrised distributions}. 
An example is the log-normal distribution $\mathcal{LN}(\mu,\sigma)$
with parameters the real number $\mu$ and the positive real number $\sigma$. 
It may be helpful to think of parameterised distributions as randomised 
functions mapping the parameters, such as $\mu$ and $\sigma$, to values in 
some range, in the case of $\mathcal{LN}(\mu,\sigma)$ the positive reals. 
As a second example, consider the simple Bernoulli (coin-flip) distribution 
$\CB(p)$ with parameter $p\in(0,1)$.
It takes value $1$ with probability $p$ and $0$ with probability $1{-}p$. 
The functions in $\Psi$ must satisfy some
technical measurability conditions to ensure that they behave well
with respect to changes of parameters. Intuitively, we want continuous
changes in the parameters to result in continuous changes of the
distribution, whatever that means technically. As examples, think of a
normal distribution $\CN(\mu,\sigma)$ and a Bernoulli distribution
$\CB(p)$.  We can compose the parameterised distributions and replace
parameters by constants to form more complex terms, but we need to
make sure that the resulting parameterised distributions still satisfy
our technical conditions. We call these \emph{$\Psi$-terms}. An
example of such a term, with two variables $s,t$, is the expression
$t+\mathcal{LN}(\ln s,1/10)$ in rule \eqref{eq:rule6}. Deterministic functions 
such as $t + s$ can be easily incorporated into the semantics as well.
A \emph{GDatalog rule} over a set $\Psi$ of parametrised distributions
is an expression
\begin{equation}
  \label{eq:4}
  R(\bar t)\gets S_1(\bar x_1),\ldots,S_m(\bar x_m), 
\end{equation}
where the \emph{body} $S_1(\bar x_1),\ldots,S_m(\bar x_m)$ is a list
of atoms over the extensional and intensional schema, just like for
normal Datalog rules, and the \emph{head} $R(\bar t)$ consists of an
intensional relation symbol $R$ and a tuple $\bar t=(t_1,\ldots,t_k)$
of ($\Psi$-)terms such that all variables of the $t_i$ appear in the
body of the rule. Of course we must make sure that the terms are of
the appropriate types, that is, the range of $t_i$ is contained in the
domain of the $i$th attribute of relation $R$. Note that in particular, all 
normal Datalog rules are GDatalog rules.

A \emph{GDatalog
  program} is a bag of GDatalog rules.

Before even touching upon the intricacies of a formal semantics
for GDatalog programs, let us explain an informal operational
semantics for GDatalog rules and programs. We have already given
the intuition in Example~\ref{exa:datalog}.  Consider the rule
\eqref{eq:4}. Let $\bar t=(t_1,\ldots,t_k)$, and let $\bar y_i$ be the
tuple of variables of the term $t_i$---we indicate this by writing
$t_i(\bar y_i)$. Note that for $\Psi$-terms, $t_i(\bar y_i)$ is a parametrised
probability distribution: if we instantiate the variables in $\bar y_i$ by
values of the appropriate type, we obtain a probability distribution.

Given an interpretation of all the body relations $S_i$ and an
assignment $\alpha$ to the variables $\bar x_i$, the rule is
applicable if for all $i$ the fact $S_i(\alpha(\bar x_i))$ holds true
under the current interpretation of $S_i$. To apply the rule, for all
$j$ we sample a value $a_j$ from the probability distributions
$t_j(\alpha(\bar y_j))$ to generate the new fact $R(a_1,\ldots,a_k)$.

We run a GDatalog program $\CG$ on a database instance over the
extensional schema in a similar way as a normal Datalog program. All
intensional relations are initialised to be empty. By repeatedly
applying the rules as described above, the program generates (random)
facts. All rule applications are stochastically independent. We
stipulate that each rule of the program (or more precisely, each
occurrence of each rule---remember a program is a bag of rules where a
rule may occur several times) can only be applied once for every
instantiation of the variables appearing in the head of the rule. The
computation terminates if no more rule can be applied, and the output
is the set of facts generated by the program during the computation,
that is, a database instance over the intensional schema. Because of
the sampling of values in the rule applications, the output is
probabilistic. We interpret it as a probabilistic database. Thus,
given a database instance over the extensional schema, a GDatalog
program generates a PDB over the intensional schema.

However, our informal description of the semantics raises several
crucial questions:
\begin{enumerate}
\item Does the program always terminate?
\item How can we be sure that the output is indeed a well-defined
  probabilistic database?
\item In which order do we apply the rules, and does this make a
  difference? 
\end{enumerate}
The answer to Question (1) is simply 'no' (in general). In
Example~\ref{exa:datalog}, we already saw that even the simple deterministic
program \eqref{eq:rule3}, \eqref{eq:rule4} may not terminate if the
input graph is cyclic. 
For probabilistic programs, the notion of termination is more complicated, 
because a program may terminate for certain random outcomes while it diverges
for other outcomes \cite{DBLP:conf/rta/BournezG05}. 
We resolve this issue by conditioning the output probability distribution 
on termination.
That is to say, a GDatalog program defines a \emph{sub-probabilistic
  database} where the probability mass over its defined space may be
smaller than $1$. 
It is an open research question to understand termination criteria for 
GDatalog programs.

We can answer Questions (2) and (3) by carefully defining a formal
semantics for GDatalog programs. 
The main results of  \cite{Grohe+2020} (in the case
of discrete probability distributions due to \cite{Barany+2017})
regarding this semantics are
(informally) summarised in the following theorem.

\begin{theorem}\label{theo:GDatalog}
  Applying the GDatalog program $\CG$ to a data{\-}base instance $D$ over 
  the extensional schema, defines a standard sub-probabilistic database 
  $\CG(D)$.

  $\CG(D)$ does not depend on the order in which the rules are
  applied, as long as the policy that is used to decide which rules to
  apply is measurable.
\end{theorem}

One final remark is that the semantics of GDatalog programs can
be lifted to probabilistic databases. That is, if we apply a GDatalog
program to a standard PDB over the extensional schema it defines a
standard sub-probabilistic database over the extensional schema.

\section{Concluding Remarks}
To enable reasoning about uncertain data with standard statistical
models, we need probabilistic databases to support continuous
probability distributions. Providing a mathematical framework for
dealing with a very general class of probability distributions, we
introduced standard PDBs \cite{GroheLindner2020}. Furthermore, we
extended Generative Datalog \cite{Barany+2017}, a declarative probabilistic
programming language for relational data, to continuous distributions,
thereby providing a flexible formalism for specifying generative
models of standard probabilistic databases.

The focus of our work was on semantical issues. Further research is
needed to address algorithmic and complexity theoretic questions.

\medskip\noindent
\textbf{Acknowledgements.}
This research is supported by the German Research Foundation (DFG)
under grant GR 1492/16-1
and
the Research Training Group 
2236 UnRAVeL.

\end{document}